\documentclass[aps,prb,showpacs,twocolumn,superscriptaddress]{revtex4}
\usepackage{epsfig,amsmath,amssymb,color}
\newcommand{\be}{\begin{equation}}
\newcommand{\ee}{\end{equation}}
\newcommand{\ra}{\rangle}
\newcommand{\la}{\langle}
\newcommand{\bit}{\begin{itemize}}
\newcommand{\eit}{\end{itemize}}
\newcommand{\bea}{\begin{eqnarray}}
\newcommand{\eea}{\end{eqnarray}}

\usepackage{epsfig}

\begin{document}
\title
{The Heisenberg
antiferromagnet on the kagome lattice with arbitrary spin:\\
A high-order coupled cluster treatment}

\author
%{O. G\"otze,$^1$ D. J. J. Farnell,$^{2}$ R. F. Bishop,$^3$ P.~H.~Y.
%~Li,$^3$ and J. Richter$^{1}$}
{O. G\"otze}
\affiliation
{Institut f\"ur Theoretische Physik, Otto-von-Guericke Universit\"at Magdeburg, 39016 Magdeburg, Germany}

\author{D. J. J. Farnell}
\affiliation
{Division of Mathematics and Statistics, 
Faculty of Advanced Technology, University of Glamorgan, 
Pontypridd CF37 1DL, Wales, United Kingdom}

\author{R. F. Bishop}
\affiliation
{School of Physics and Astronomy, The University of Manchester,
Manchester M13 9PL, UK}

\author{P.~H.~Y.~Li}
\affiliation
{School of Physics and Astronomy, The University of Manchester,
Manchester M13 9PL, UK}

\author{J. Richter}
\affiliation
{Institut f\"ur Theoretische Physik, Otto-von-Guericke Universit\"at Magdeburg, 39016 Magdeburg, Germany}

\begin{abstract}
Starting with the
$\sqrt{3}\times\sqrt{3}$  and the
$q=0$ states as reference states
we use the coupled cluster method to high orders of approximation
to investigate  the ground state of the 
Heisenberg antiferromagnet   on the kagome lattice for spin quantum numbers
$s=1/2,1,3/2,2,5/2$, and $3$.
Our data for the ground-state energy for $s=1/2$ are in good agreement 
with recent 
large-scale  density-matrix renormalization group  and exact
diagonalization data.
We find that  the ground-state selection depends on the spin quantum number  $s$.
While for the extreme quantum case, $s=1/2$,
the $q=0$ state is energetically favored by quantum fluctuations, for any
$s>1/2$ the $\sqrt{3}\times\sqrt{3}$ state is selected.
\textcolor{black}{
For both the
$\sqrt{3}\times\sqrt{3}$  and the
$q=0$ states the magnetic order is strongly suppressed by quantum
fluctuations. Within our coupled cluster method we 
get vanishing values for the order parameter (sublattice magnetization) $M$ for
$s=1/2$ and $s=1$, but (small) nonzero values for $M$ for $s>1$.
Using the data for the  ground-state energy and the order parameter for
$s=3/2,2,5/2$, and $3$ we also estimate the leading  
quantum corrections to the classical values.}
\end{abstract}
% PACS, the Physics and Astronomy Classification
                             % Scheme. 
\maketitle

\section{Introduction}
The investigation of the low-energy physics of the Heisenberg antiferromagnet
(HAFM) 
\be
H = \sum_{<i,j>}{\bf s}_i\cdot{\bf  s}_j
\label{H}
\ee
on the kagome lattice is one of the most
challenging problems in the field of frustrated quantum magnetism.
The sum over $\langle i,j \rangle$ runs over all nearest-neighbor pairs of sites on the lattice, counting each bond once only, and ${\bf s}_{i} \equiv (s^{x}_{i},s^{y}_{i},s^{z}_{i})$ is the spin operator on site $i$.  Although, there has been an intensive discussion of the problem over many years
applying various theoretical methods, (and see,
e.g.,~Refs.~\onlinecite{marston91,Harris1992,Chalker1992,huse1992,sachdev1992,singh1992,chub92,Leung1993,suzuki1994,Waldtmann1998,Zeng1995,
henley1995,Mambrini2000,farnell2001,bern2002,Nikolic2003,schmal2004,Budnik2004,Capponi2004,Sindzingre2009,Singh2007,Singh2008,
Jiang2008,henley2009,Poilblanc2010,Bishop2010,Evenbly2010,Yan2011,lauchli2011,lee2011,nakano2011,tay2011,cepas2011,iqbal2011}), 
no conclusive answer on the nature of
the ground state (GS) and the existence of a spin gap has been found.

While for many years a spin-liquid GS was
favored,\cite{Chalker1992,Zeng1995,Waldtmann1998} recently        
arguments have been given for a valence-bond crystal GS with a large unit cell
of 36 sites that breaks
the symmetry of the underlying kagome lattice.\cite{Nikolic2003,Singh2007,Singh2008}
However, very recently this  valence-bond picture has been rechecked by
large-scale numerics\cite{Yan2011,lauchli2011,nakano2011} and once again the 
spin-liquid GS is favored.

Although, large-scale  density-matrix renormalization group (DMRG) and exact
diagonalization (ED) calculations seem to be most effective to study the
low-energy physics of the kagome HAFM,
complementary methods (and see, e.g., Refs.~\onlinecite{Evenbly2010,lee2011,tay2011,iqbal2011}),
are highly desirable to shed further light on the challenging problem.

A method which has been successfully applied to strongly frustrated quantum
magnets is the coupled cluster method (CCM) (and see, e.g.,
Refs.~\onlinecite{farnell2001,Bishop2010,rachid05,Schm:2006,
rachid08,bishop08,farnell09,richter2010,farnell11,farnell11a}).
In the present paper we apply the CCM in high orders of approximation to the kagome HAFM.

\begin{figure}[ht]
\begin{center}
\epsfig{file=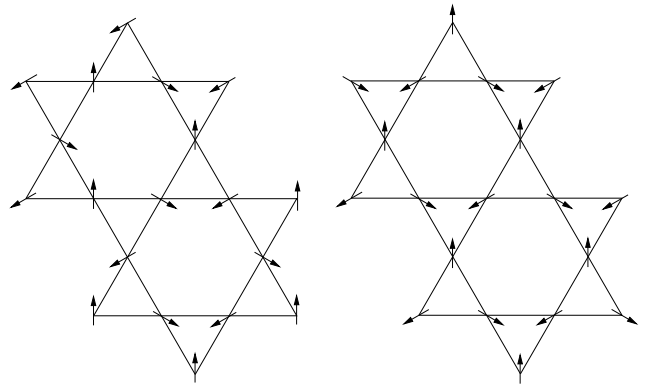,scale=0.45,angle=0.0}
\end{center}
\caption{
Illustration of the $\sqrt{3}\times\sqrt{3}$ (left) and the
$q=0$ (right) classical GS  of the kagome HAFM. 
}
\label{fig1}
\end{figure}

\section{Coupled cluster method}
For the sake of brevity we illustrate here only some relevant features of
the coupled cluster method (CCM). 
For more general information on the methodology of the CCM, see, e.g.,
Refs.~\onlinecite{zeng98,bishop98a,bishop99,bishop00,farnell02,bishop04}.

We first mention that the CCM approach yields results directly in 
the thermodynamic limit $N\to\infty$, 
where $N$ is the number of lattice sites (and hence spins).  
The starting point for a CCM
calculation is the choice of a normalized reference  state
$|\Phi\rangle$ that is typically a classical GS of the model.  
For the kagome HAFM
we choose the  $\sqrt{3}\times\sqrt{3}$ and the
$q=0$ states illustrated in Fig. \ref{fig1}, (and see also, 
e.g., Refs.~\onlinecite{Harris1992,chub92,tay2011} for further details).
Then we perform a rotation of the local axes of each of 
the spins such that all spins in the reference state align along the
negative $z$ axis.  
In this new set of local spin coordinates 
a complete set of 
mutually commuting multispin
creation operators $C_I^+ \equiv (C^{-}_{I})^{\dagger}$ related to this reference state is defined by
\begin{equation}
\label{set1} |{\Phi}\ra = |\downarrow\downarrow\downarrow\cdots\rangle ; \mbox{ }
C_I^+ 
= { s}_{n}^+ \, , \, { s}_{n}^+{ s}_{m}^+ \, , \, { s}_{n}^+{ s}_{m}^+{
s}_{k}^+ \, , \, \ldots \; ,
\end{equation}
where $s^{+}_{n} \equiv s^{x}_{n} + is^{y}_{n}$, the indices $n,m,k,\ldots$ denote arbitrary lattice sites, and the
components of the spin operators are defined 
in the local rotated coordinate frames. 
Note that for spins of quantum number $s$, each site index in each configuration index $I$ in
Eq.~({\ref{set1}) can be repeated up to a maximum of $2s$ times.
With the set $\{|\Phi\rangle, C_I^+\}$ thus defined, the CCM parametrizations of 
the ket 
and bra GS eigenvectors
$|\Psi\ra$ 
and $\la\tilde{\Psi}|$ 
of the spin system 
are given  by
\begin{eqnarray}
\label{eq5} 
|\Psi\ra=e^S|\Phi\ra \; , \mbox{ } S=\sum_{I\neq 0}a_IC_I^+ \; \\
\label{eq5b}
\la \tilde{ \Psi}|=\la \Phi |\tilde{S}e^{-S} \; , \mbox{ } \tilde{S}=1+
\sum_{I\neq 0}\tilde{a}_IC_I^{-} \; .
\end{eqnarray}
We have defined $C_{0}^{+} \equiv 1$, and the normalization of the states is clearly such that $\langle \tilde{ \Psi}| \Psi \rangle = 
\langle \Phi| \Psi \rangle = \langle \Phi| \Phi \rangle \equiv 1$. 
The CCM correlation operators, $S$ and $\tilde{S}$, contain the correlation coefficients,
$a_I$ and $\tilde{a}_I$, 
which can be determined by the CCM ket-state 
and bra-state
equations
\begin{eqnarray}
\label{eq6}
\langle\Phi|C_I^-e^{-S}He^S|\Phi\rangle = 0 \;\; ; \; \forall I\neq 0  \\ 
\label{eq6a}\langle\Phi|{\tilde S}e^{-S}[H, C_I^+]e^S|\Phi\rangle = 0 \; \; ; \; \forall
I\neq 0.
\end{eqnarray}
Equations (\ref{eq6}) and (\ref{eq6a}) are fully equivalent to the GS Schr\"{o}dinger equations for the ket and bra states.  They follow readily from the requirement that the GS energy functional $\langle\tilde{\Psi}|H|\Psi\rangle$ be stationary with respect to variations in all of the correlation coefficients $\tilde{a}_{I}$ and $a_{I}$ respectively ($\forall I \neq 0)$.
Each ket-state 
or bra-state 
equation belongs to a certain configuration index $I$,
i.e., it corresponds to a certain set (configuration) of lattice sites
$n,m,k,\dots\;$, as in Eq.~(\ref{set1}).
Using the Schr\"odinger equation, $H|\Psi\ra=E|\Psi\ra$, we can now write
the GS energy as $E=\la\Phi|e^{-S}He^S|\Phi\ra$.
The magnetic order parameter (sublattice magnetization) is given
by $ M = -\frac{1}{N} \sum_{i=1}^N \la\tilde\Psi|{ s}_i^z|\Psi\ra$, where
${s}_i^z$
is expressed in the transformed coordinate system, and $N(\rightarrow \infty)$ is the number of lattice sites.. 

If we would be able to consider all creation and annihilation operators
$C_I^+$  and $C_I^-$, 
i.e. all sets (configurations)  of lattice sites, in the 
CCM correlation operators $S$ and $\tilde{S}$ we would get, in principle, the exact
eigenstate.\cite{bishop98a}
However, for the many-body quantum system under consideration
it is necessary to use approximation 
schemes in order to truncate the expansions of $S$ 
and $\tilde S$ 
in  Eqs.~(\ref{eq5}) and (\ref{eq5b}) 
in a practical calculation.
Then the approximate results for the GS energy $E$ and the order
parameter $M$ will depend certainly on the choice of the reference state.

We use for spin quantum number $s=1/2$ the so-called LSUB$n$
approximation scheme to truncate the expansions 
of $S$ and $\tilde S$ 
in Eqs.~(\ref{eq5}) and (\ref{eq5b}), where we 
include only $n$ or fewer correlated spins in all configurations (or lattice animals in the language of graph theory) which 
span a range of no more than $n$ contiguous 
lattice  sites, where a set of sites is defined to be contiguous if every site has at least one other in the set as a nearest neighbor (and for more details see Refs.~\onlinecite{zeng98,bishop00,farnell02,bishop04}). 
Using an efficient 
parallelized CCM code \cite{cccm} we are able to solve the CCM equations up
to LSUB10 for $s=1/2$ (where, e.g. for the $q=0$
reference state a set of 238010 coupled ket-state equations has to be
solved), which goes significantly beyond earlier CCM
calculations for the
kagome HAFM.\cite{farnell2001,Bishop2010} %$\sqrt{3}\times\sqrt{3}$ Lsub6 in farnell2001 und q=0 LSUB8 in Bishop2010
Moreover, we also use the CCM to
consider spin quantum numbers $s>1/2$.

Since the LSUB$n$ approximation becomes exact for $n \to \infty$ (as also so does the alternative SUB$n$-$n$ scheme that we introduce and use below for values of the spin quantum number $s>1/2$ in 
Sec.~\ref{larger_S}), it is useful 
to extrapolate the `raw' LSUB$n$ (or SUB$n$-$n$)
data to the limit $n \to \infty$. 
There is ample experience regarding how one should
extrapolate the GS energy per site $e_0(n) \equiv E(n)/N$ and the magnetic order parameter
$M(n)$. For the GS energy per spin $e_0(n) = a_0 + a_1(1/n)^2 + a_2(1/n)^4$ is a very
well-tested extrapolation ansatz.\cite{bishop00,bishop04,rachid05,
Schm:2006,bishop08,rachid08,richter2010}  An appropriate extrapolation rule
for the magnetic order parameter of highly frustrated systems
is\cite{bishop08,rachid08,richter2010}
$M(n)=b_0+b_1(1/n)^{1/2}+b_2(1/n)^{3/2}$.
Moreover, we know from Refs.~\onlinecite{bishop08,rachid08,richter2010} that low
levels
of approximation conform poorly to these
rules. Hence, we
exclude the $n=2$ and $n=3$  data from the extrapolations.

For the solution of the CCM equations we rewrite the Hamiltonian (\ref{H})  
in the rotated coordination frame of the local quantization axis
\be\begin{aligned}
H_\lambda&=\sum_{<i\rightarrow j>}\biggl[-\frac{1}{2}\left(\lambda s^{x}_{i}s^{x}_{j} 
+ s^{z}_{i}s^{z}_{j}\right) + \lambda s^{y}_{i}s^{y}_{j}\\
&+ \frac{\sqrt{3}}{2}\lambda\left(-s^{x}_{i}s^{z}_{j} +
s^{z}_{i}s^{x}_{j}\right)\biggr],
\end{aligned}\label{Hlambda}
\ee
where we have further introduced an anisotropy parameter $\lambda$ that now multiplies the 
non-Ising terms, similar to what was done in Refs.~\onlinecite{singh1992,bishop99}. 
Note that the symbol $<i\rightarrow j>$ on the sum in Eq. (\ref{Hlambda}) now indicates a directionality for the 
nearest-neighbor bonds, (and see, e.g., Refs.~\onlinecite{singh1992,bishop99,farnell09}),
which is different for the  $\sqrt{3}\times\sqrt{3}$ and the
$q=0$ reference states.    
Starting at $\lambda=0$, where the corresponding reference states are
eigenstates of $H_\lambda$, we can slowly increase $\lambda$ and hence trace the CCM solutions out to the true
kagome point at $\lambda=1$.
Moreover, $\lambda$ can be understood as a parameter that tunes the strength of the quantum
fluctuations.

\begin{figure}[ht]
\begin{center}
\epsfig{file=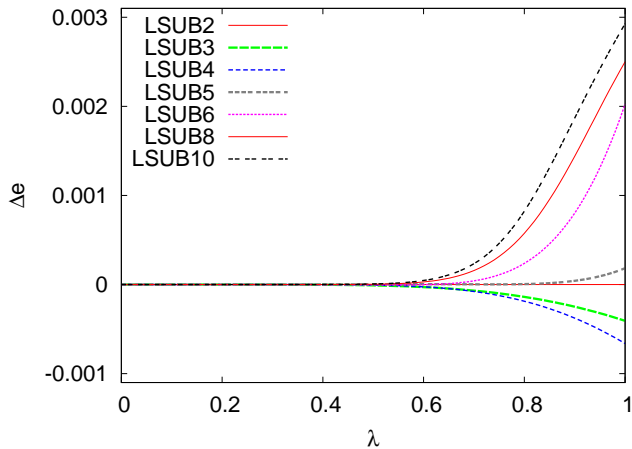,scale=0.7,angle=0.0}
\end{center}
\caption{Difference of GS energies per site, $\Delta e \equiv
e_0^{\sqrt{3}\times\sqrt{3}} - e_0^{q=0}$, between the 
$\sqrt{3}\times\sqrt{3}$ and $q=0$ states of the spin-$1/2$ kagome HAFM, for various CCM LSUB$n$
approximations and spin quantum number $s=1/2$.
}
\label{fig2}
\end{figure}

\begin{figure}[ht]
\begin{center}
\epsfig{file=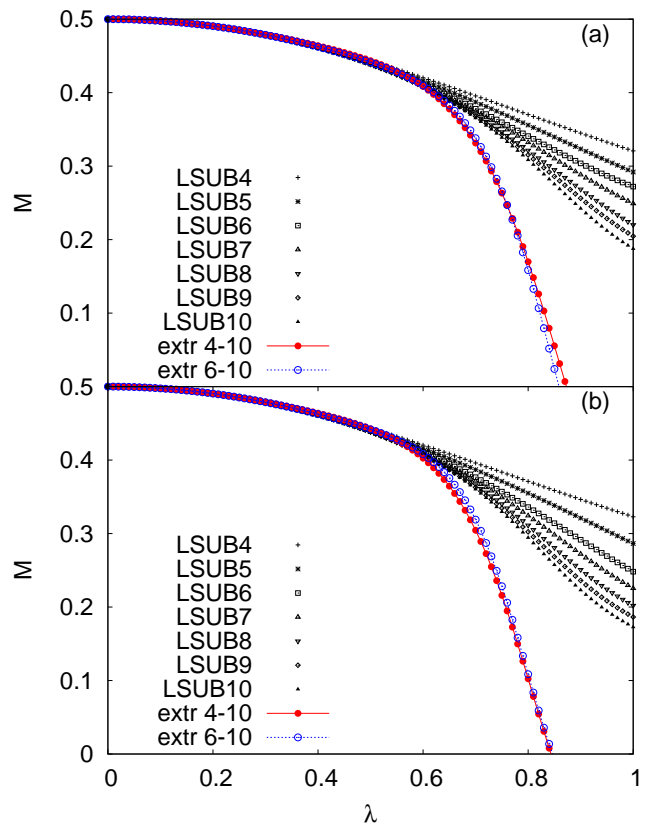,scale=0.55,angle=0.0}
\end{center}
\caption{CCM-LSUB$n$ data for the magnetic order parameter $M$  versus $\lambda$ for the spin-$1/2$ kagome HAFM, 
for (a) the  $\sqrt{3}\times\sqrt{3}$ reference state and (b) the $q=0$ reference state.
For the extrapolations to $n \to \infty$  according to 
$M(n)=b_0+b_1(1/n)^{1/2}+b_2(1/n)^{3/2}$
we have used  LSUB$n$ data for $n=4,5,\ldots,10$ as well as for
$n=6,7,\ldots,10$. 
}
\label{fig3}
\end{figure}

\section{Results}

\subsection{The extreme quantum case:  $s=1/2$}

\begin{table}
\caption{CCM results for the spin-$1/2$ HAFM on the kagome lattice (i.e., at
$\lambda =
1$).  The quantity $e_0 \equiv E/N$ is the GS energy per spin 
and $M$ is the magnetic order
parameter (sublattice magnetization).
The LSUB$n$ results are extrapolated  to $n\to \infty$ according
to 
 $e_0(n) = a_0 + a_1(1/n)^2 + a_2(1/n)^4$
and
$M(n)=b_0+b_1(1/n)^{1/2}+b_2(1/n)^{3/2}$
using  LSUB$n$ data for $n=4,5,\ldots,10$ as well as for
$n=6,7,\ldots,10$.
}         
\begin{tabular}{@{}|c|c|c|@{}}\hline 
{\bf $\sqrt{3}\times\sqrt{3}$}      &$e_0$             &$M$              \\\hline
LSUB4                               &-0.408728         &0.320702           \\\hline
LSUB5                               &-0.414235         &0.291917           \\\hline
LSUB6                               &-0.418052         &0.272109           \\\hline
LSUB7                               &-0.420677         &0.248989           \\\hline
LSUB8                               &-0.423554         &0.219994           \\\hline
LSUB9                               &-0.424962         &0.204661           \\\hline
LSUB10                              &-0.426485         &0.187634           \\\hline
{\bf Extrapolated (4-10)}           &{\bf  -0.4318 }   &$<0$               \\\hline
%{\bf Extrapolated (4,6,8,10)}       &{\bf  -0.4323 }   &$<0$               \\\hline
{\bf Extrapolated (6-10)}           &{\bf  -0.4336 }   &$<0$               \\\hline
%{\bf Extrapolated (6,8,10)}         &{\bf  -0.4323 }   &$<0$               \\\hline
\hline
{\bf $q=0$}                         &$e_0$             &$M$              \\\hline
LSUB4                               &-0.408066         &0.322860           \\\hline
LSUB5                               &-0.414418         &0.286462           \\\hline
LSUB6                               &-0.420078         &0.248078           \\\hline
LSUB7                               &-0.423126         &0.225356           \\\hline
LSUB8                               &-0.426054         &0.202074           \\\hline
LSUB9                               &-0.427952         &0.186435           \\\hline
LSUB10                              &-0.429413         &0.172742           \\\hline
{\bf Extrapolated (4-10)}           &{\bf  -0.4357 }   &$<0$               \\\hline
%{\bf Extrapolated (4,6,8,10)}       &{\bf  -0.4355 }   &$<0$               \\\hline
{\bf Extrapolated (6-10)}           &{\bf  -0.4372 }   &$<0$               \\\hline
%{\bf Extrapolated (6,8,10)}         &{\bf  -0.4363 }   &$<0$               \\\hline
\hline
\multicolumn{3}{|c|}{\bf other recent results}\\\hline
Ref.~\onlinecite{Singh2007}		    &-0.433            & -                 \\\hline
Ref.~\onlinecite{Evenbly2010} 		    &-0.4322           & -                 \\\hline
%DMRG u.b.       		    &-0.4332           & -                 \\\hline
Ref.~\onlinecite{lauchli2011},  $N=42$ (type a)     &-0.437999          & -                 \\\hline
Ref.~\onlinecite{lauchli2011},  $N=42$ (type b)    &-0.438143            & -                 \\\hline
Ref.~\onlinecite{Yan2011}                   &-0.4379           & -                 \\\hline
\hline
\end{tabular}
\label{table1}
\end{table}

We start with the CCM investigation  of the kagome HAFM for spin quantum number
$s=1/2$. 
%As already mentioned above in the limit $n \to \infty$ the result
%should not depend on the choice of the reference state.
%However, in 
At a given finite level of the CCM LSUB$n$ scheme the treatment
of quantum
effects is performed in an approximate manner. 
Certainly, the treatment of
quantum effects becomes better as the level of approximation $n$ is increased.
In previous studies of the GS selection based on an expansion around the
classical limit\cite{chub92,sachdev1992,henley1995} the $\sqrt{3}\times\sqrt{3}$ state
was found to be selected by quantum fluctuations.
We present  our results for the GS energy per site
in Fig.~\ref{fig2}, where the dependence on the anisotropy
parameter $\lambda$ of the 
difference in the energies per site between the two states considered, $\Delta e \equiv e_0^{\sqrt{3}\times\sqrt{3}}
 - e_0^{q=0}$,
is shown for various LSUB$n$ approximations.
%The difference in the two variants of the extrapolation (including or excluding
%LSUB4 and LSUB5) may be considered as an error bar for the extrapolated
%energy.  
Interestingly, the GS selection depends on the LSUB$n$ truncation index $n$.
Just as in linear spin-wave theory,\cite{Harris1992,henley1995}
there is also no GS selection (i.e., $\Delta e=0$) 
at the CCM-LSUB$2$ level, thereby indicating a poor consideration of quantum effects at the lowest LSUB$n$ order.
As the level of approximation $n$ is increased  
we first find that $\Delta e<0$ for $n=3$ and $n=4$ (i.e., 
the $\sqrt{3}\times\sqrt{3}$ state is selected in accordance with
previous findings\cite{chub92,henley1995}), but as $n$ is further increased we
then find that $\Delta e>0$ for $n>4$
(i.e., the  $q=0$ state is selected).
Bearing in mind that quantum effects are better taken into account at higher LSUB$n$ levels we might argue that strong quantum fluctuations indeed favor the $q=0$ state.
Note that this line of argument is also supported by our CCM results below for spin quantum
numbers $s> 1/2$ (i.e., generally speaking, where quantum fluctuations are weaker), 
where in all levels of approximations the $\sqrt{3}\times\sqrt{3}$ state is
selected (and see our discussion below in Sec.~\ref{larger_S}).

In Fig.~\ref{fig3} we show the magnetic order parameter as a function of the anisotropy parameter $\lambda$. At $\lambda=0$ we have 
$M=s=1/2$, since the corresponding reference state is the exact GS of
$H_{\lambda=0}$. As $\lambda$ is increased the order parameter
decreases
monotonically. At a certain value of $\lambda$, near the true kagome point
$\lambda=1$, the extrapolated order parameter vanishes, thus  indicating that the GS
is magnetically disordered. The difference in the two variants of the
extrapolation (including or excluding
LSUB4 and LSUB5) may be considered as an error bar for the extrapolated
order parameter.

Next we analyze the model for $\lambda=1$ in more detail. 
The CCM-LSUB$n$ data, as well as the extrapolated data, are listed in
Table~\ref{table1}. Moreover, we present results for the GS energy
obtained by other methods for comparison. While the values $e_0=-0.4322$ obtained in 
Ref.~\onlinecite{Evenbly2010} and $e_0=-0.4332$ obtained in
Ref.~\onlinecite{Yan2011} can be considered as rigorous  upper bounds of the GS
energy, the large-scale DMRG
result $e_0=-0.4379$ obtained in
Ref.~\onlinecite{Yan2011} seems to be the most accurate estimate presently available.
The lowest extrapolated CCM energy is $e_0=-0.4372$ obtained for the $q=0$ reference state
using CCM-LSUB$n$ results for $n=6,7,8,9,10$ for the extrapolation.
This CCM estimate is very close to the DMRG result of
Ref.~\onlinecite{Yan2011}.

\begin{figure}[ht]
\begin{center}
\epsfig{file=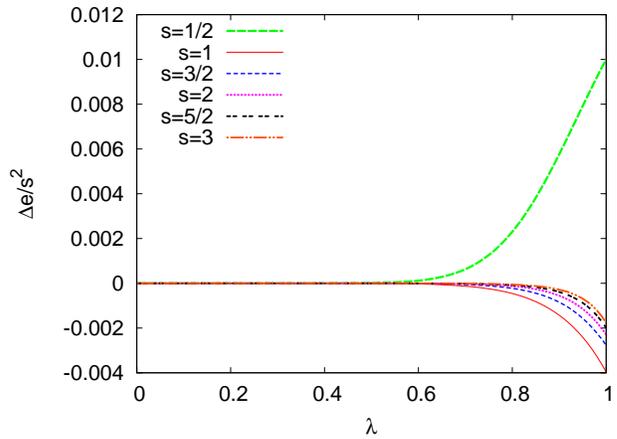,scale=0.67,angle=0.0}
\end{center}
\caption{
Difference of the GS energies per site, $\Delta e \equiv 
e_0^{\sqrt{3}\times\sqrt{3}} - e_0^{q=0}$, 
between the $\sqrt{3}\times\sqrt{3}$ and $q=0$ states
of the kagome HAFM, calculated for the CCM SUB$8$-$8$
approximation and for spin quantum numbers $s=1/2,1,3/2,2,5/2,3$. 
}
\label{fig4}
\end{figure}

\begin{figure}[ht]
\begin{center}
\epsfig{file=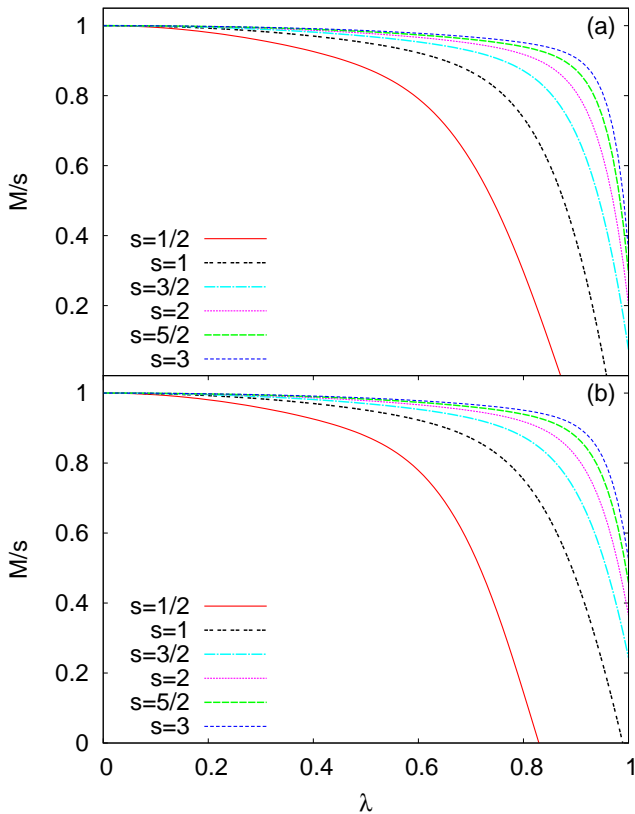,scale=0.55,angle=0.0}
\end{center}
\caption{
Extrapolated magnetic order parameter, $M/s$, versus $\lambda$ for (a) the $\sqrt{3}\times\sqrt{3}$ reference state and (b) the $q=0$ reference state of the kagome HAFM, for various values of the 
spin quantum number $s$.
For the extrapolations to $n \to \infty$  according to
$M(n)=b_0+b_1(1/n)^{1/2}+b_2(1/n)^{3/2}$
we have used  SUB$n$-$n$ data for $n=4,5,\ldots,8$. [Note that even for
the $s=1/2$ case we have excluded the (available) LSUB9 and LSUB10 data
here to be consistent with the $s>1/2$ cases.] 
}
\label{fig5}
\end{figure}

\subsection{Higher spin quantum numbers:  $s>1/2$}
\label{larger_S}

Although several magnetic kagome compounds carry spins with $s>1/2$, such as  
the $s=3/2$ magnet KCr$_3$(OH)$_6$(SO$_4$)$_2$,\cite{comp1} or the $s=5/2$
compound (H$_3$O)Fe$_3$(OH)$_6$(SO$_4$)$_2$,\cite{comp2}    
far fewer theoretical results are available for those higher spin quantum numbers.
In the classical limit, $s \to \infty$, thermal fluctuations may lead to 
$\sqrt{3}\times\sqrt{3}$ long-range order as $T \to
0$.\cite{huse1992,henley2009} 
In most papers dealing with large-spin quantum models it has been found that quantum fluctuations select the
$\sqrt{3}\times\sqrt{3}$
state.\cite{chub92,sachdev1992,henley1995,cepas2011}
Moreover, magnetic long-range order might be possible  for
higher  spin values.\cite{sachdev1992,cepas2011}

For  our CCM approach for $s>1/2$ we use (instead of the LSUB$n$ scheme) the alternative SUB$n$-$m$ 
approximation scheme to truncate the expansions of 
$S$ and $\tilde S$ 
in Eqs.~(\ref{eq5}) and (\ref{eq5b}).  This is because as $s$ increases the number of fundamental configuration $I$ 
retained at a given LSUB$n$ level also increases, 
since each spin at any site $i$  may be raised up to $2s$ times by its raising operator $s^{+}_{i}$, 
and hence each site index $i$ may be repeated up to $2s$ 
times in the operators $C^{+}_{I}$ of Eq.~(\ref{set1}). In the SUB$n$-$m$ scheme we include 
no more than $n$ spin flips spanning a range of no more than 
$m$ contiguous lattice sites.\cite{farnell02,bishop04}
In what follows we consider the case $n=m$, i.e. SUB$n$-$n$, which for $s=1/2$ is identical to the LSUB$n$ scheme.
Since the number of coupled ket-state equations for a certain level of 
SUB$n$-$n$ approximation increases with increasing spin quantum number $s$,
the highest level of approximation we can consider is SUB8-8 for
$s=1,3/2,2,5/2,3$.
The maximum number of ket-state equations we have to take into account
is  416126 for $s=3$.
For the extrapolation to $n \to \infty$ we use the same extrapolation
formulas as for $s=1/2$, and consider the SUB$n$-$n$ data for $n=4,5,6,7,8$.

First we discuss the GS selection. In Fig.~\ref{fig4}
we present the dependence on the anisotropy
parameter $\lambda$ of the energy
difference per site, $\Delta e \equiv e_0^{\sqrt{3}\times\sqrt{3}} - e_0^{q=0}$,
for the highest level of approximation that we have performed 
(viz., SUB$8$-$8$), and for values of the spin quantum number 
$s=1/2,1,\ldots,3$.
We find that the $\sqrt{3}\times\sqrt{3}$ state
is selected for all values $s>1/2$.
This is in agreement with previous studies based on
an expansion around the
classical limit,\cite{chub92,sachdev1992,henley1995} but it is in contrast to our
findings for the extreme quantum case $s=1/2$. Hence, interestingly, our results suggest
that for the frustrated quantum spin system under consideration the  $s=1/2$
case and the cases $s>1/2$ may  exhibit different behavior.  It is 
interesting to note that a similar effect has also been observed for a frustrated quantum spin chain.\cite{krivnov07}
We see clearly that the energy
difference $\Delta e$ scaled by $s^2$ decreases monotonically with increasing
$s$, thereby demonstrating that for $s \to \infty$ the  $\sqrt{3}\times\sqrt{3}$ and the
$q=0$ states become degenerate.

\begin{table}[htb]
    \centering
\caption{CCM results for the HAFM on the kagome lattice (i.e., at
$\lambda =
1$) for spin quantum numbers $s=1,3/2,2,5/2,3$. 
The quantity $e_0 \equiv E/N$ is the GS energy per spin and $M$ is the magnetic order
parameter (sublattice magnetization).
The SUB$n$-$n$ results are extrapolated  to $n\to \infty$ according
to
 $e_0(n) = a_0 + a_1(1/n)^2 + a_2(1/n)^4$
and
$M(n)=b_0+b_1(1/n)^{1/2}+b_2(1/n)^{3/2}$
using  SUB$n$-$n$ data for $n=4,5,6,7,8$.
}
    \begin{tabular}{|c|rr|rr|c}\hline\hline
    \parbox[0pt][1.5em][c]{0cm}{}        & \multicolumn{2}{c|}{$\sqrt{3}\times\sqrt{3}$} & \multicolumn{2}{c|}{$q=0$} \\\cline{2-5}\hline\hline
    $s=1$ & $e_0/s^2$ & $M/s$& $e_0/s^2$ & $M/s$\\\hline
%                    q=sqrt(3)            q=0
    SUB8-8            & -1.383644 &  0.580079 & -1.379680 &  0.607293    \\
    extr4-8           & -1.4031 & $< 0$       & -1.3965 &  $< 0$    \\\hline\hline
    $s=3/2$ & $e_0/s^2$ & $M/s$& $e_0/s^2$ & $M/s$\\\hline
%                    q=sqrt(3)            q=0
    SUB8-8            & -1.257354 &  0.690229 & -1.254588 &  0.709167    \\\hline
    extr4-8           & -1.2680 &  0.0744 & -1.2643 &  0.2438    \\\hline\hline
    $s=2$ & $e_0/s^2$ & $M/s$& $e_0/s^2$ & $M/s$\\\hline
%                    q=sqrt(3)            q=0
    SUB8-8            & -1.195442 &  0.735642 & -1.193145 &  0.754580    \\\hline
    extr4-8           & -1.2026 &  0.2029 & -1.2000 &  0.3645    \\\hline\hline
    $s=5/2$ & $e_0/s^2$& $M/s$& $e_0/s^2$ & $M/s$\\\hline
%                    q=sqrt(3)            q=0
    SUB8-8            & -1.157697 &  0.766290 & -1.155703 &  0.785822    \\\hline
    extr4-8           & -1.1627 &  0.2942 & -1.1607 &  0.4586    \\\hline\hline
    $s=3$ & $e_0/s^2$ & $M/s$& $e_0/s^2$ & $M/s$\\\hline
%                    q=sqrt(3)            q=0
    SUB8-8            &-1.132263 &  0.788722 &-1.130497 &  0.808862    \\\hline
    extr4-8           &-1.1360 &  0.3583 &-1.1344 &  0.5256    \\\hline\hline
    $s \to \infty $ & $e_0/s^2$ & $M/s$& $e_0/s^2$ & $M/s$\\\hline
%                    q=sqrt(3)            q=0
    exact            &-1 &  1 &-1 & 1           \\\hline\hline
    \end{tabular}
\label{table2}
\end{table}

Next we discuss the magnetic order parameter $M$. 
To compare results for various values of $s$ it is useful to
consider $M/s$.
Since for $s \to \infty $
the chosen reference state is an eigenstate  we would get $M/s=1$ within our CCM approach in this classical limit. 
Hence, we may expect that by
increasing the spin quantum number the quantity $M/s$ becomes nonzero  for a certain value $s > s_0$ in the whole range $0 \le \lambda \le 1$ (and see also, e.g., Ref.~\onlinecite{sachdev1992}). 
In Fig.~\ref{fig5} 
we show the extrapolated magnetic order parameter for both reference states as a function of the
anisotropy parameter $\lambda$, for values $s=1/2,1,\cdots,5/2,3$.
For large values of $s$ the scaled order parameter becomes almost constant, $M/s \sim 1$, over a wide range of $\lambda$ values.  However, as the true kagome
point at $\lambda=1$ is approached we find a steep decay of $M/s$.  Nevertheless, only for $s=1/2$ (as discussed already above)
and for $s=1$ does the extrapolated order parameter vanish at $\lambda=1$, whereas $M/s$
remains nonzero for $s>1$. Hence, our data suggest that for higher values of
$s$ a $\sqrt{3}\times\sqrt{3}$ magnetic order might be possible. 
To provide more detailed information on the GS properties of the kagome HAFM
we present in Table~\ref{table2} 
CCM-SUB$8$-$8$ data as well as extrapolated data at $\lambda=1$, for
values of the spin quantum number $s=1,3/2,3,5/2,3$.
We see clearly that the scaled GS energy per spin approaches the classical value, $e_0/s^2=-1$, quite rapidly as $s$ is increased.
On the other hand, even for the largest spin considered here (viz., $s=3$)    
the extrapolated order parameter remains relatively small, particularly for the    
$\sqrt{3}\times\sqrt{3}$ state.

\begin{figure}[ht]
\begin{center}
\epsfig{file=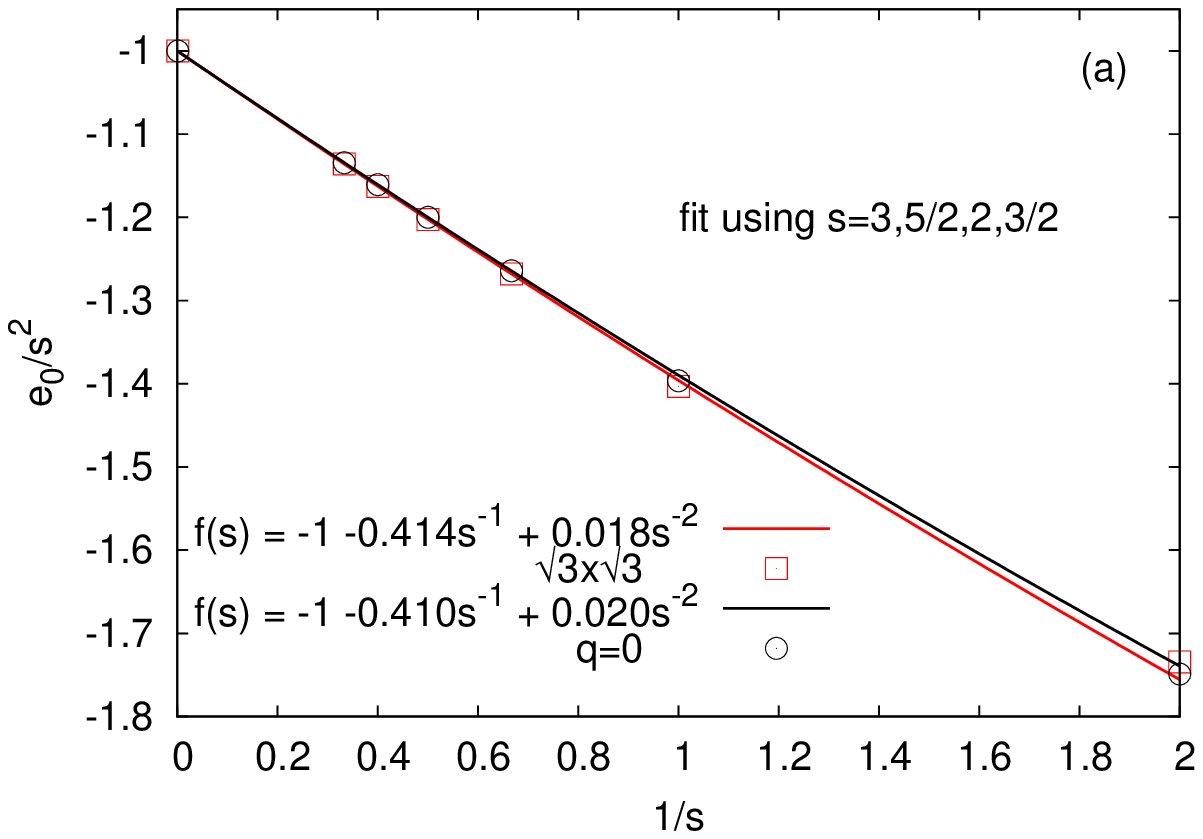,scale=0.65,angle=0.0}
\epsfig{file=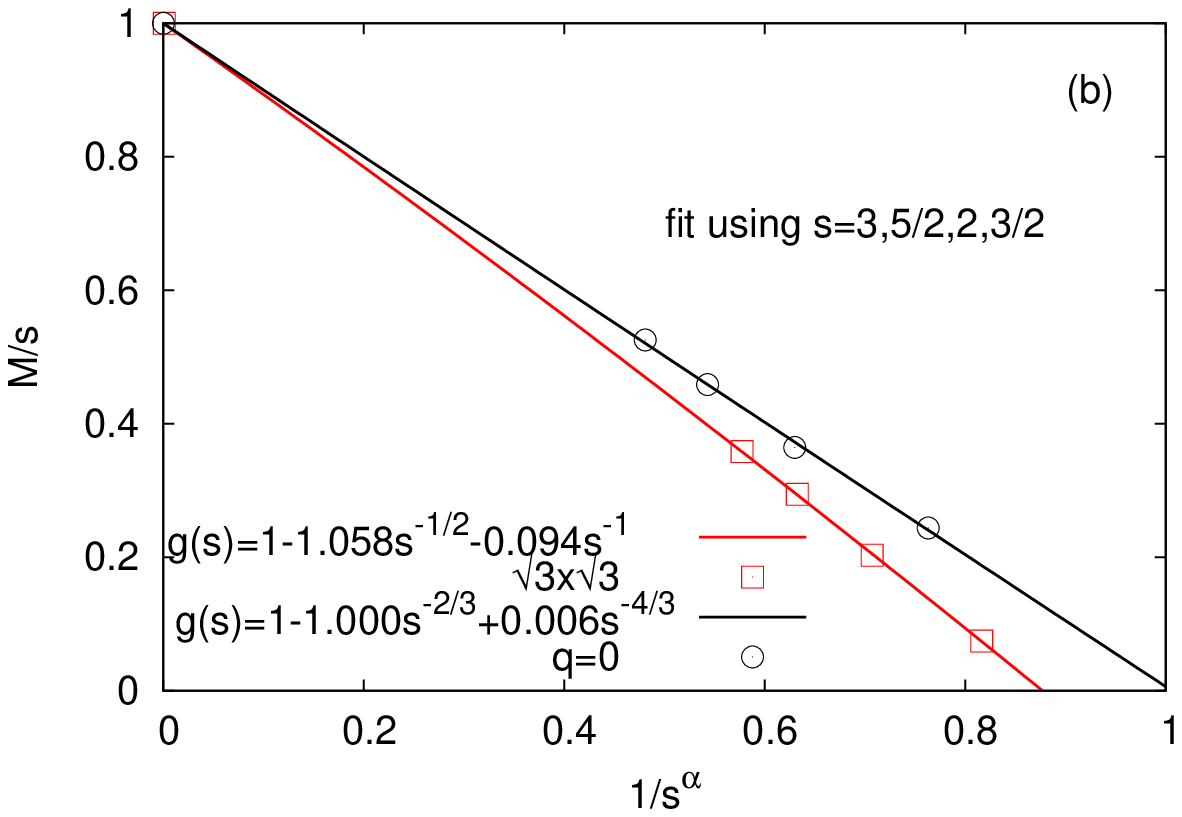,scale=0.65,angle=0.0}
\end{center}
\caption{Dependence on the spin quantum number $s$ of (a) the extrapolated scaled GS energy per spin, $e_0/s^2$, and  
(b) the extrapolated scaled magnetic order parameter, $M/s$, of the kagome HAFM.  For $M/s$ the exponent $\alpha = 1/2$ for the $\sqrt{3}\times\sqrt{3}$ state and $\alpha=2/3$ for the $q=0$
state.
(symbols: CCM data points, lines fits to the data points).
}
\label{fig6}
\end{figure}

Our CCM data for $e_0$ and $M$, available up to $s=3$, together with the known
results in the classical limit, $\lim_{s \to \infty} e_0/s^2 = -1$ and $\lim_{s \to
\infty}
M/s = 1$, also allow us to discuss the $s$-dependence of $e_0/s^2$ and $M/s$ in
the large-$s$ limit.
In spin-wave theories one typically obtains expansions for $e_{o}$ and $M$ in powers of $1/s$.\cite{oitmaa1992,chub94}
For the GS energy of the kagome HAFM the standard linear spin-wave theory yields for both the $\sqrt{3}\times\sqrt{3}$ and the $q=0$ states, $e_0/s^2=-1-0.4412/s$ (and see Refs.~\onlinecite{Harris1992,suzuki1994}).  On the other hand, 
due to the presence of the flat zero mode in the kagome HAFM, the integral for
the order parameter diverges.\cite{suzuki1994,star2004}  
Using an effective spin-wave theory, in which short-wavelength fluctuations are neglected, Asakawa and
Suzuki\cite{suzuki1994} obtained     
$M/s= 1 - 0.336/s$, whereas Chubukov\cite{chub92} found fluctuation corrections
proportional to $s^{-2/3}$ (in contrast to conventional spin-waves) by using
a self-consistent spin-wave approach.

Here we use our extrapolated CCM data for $s=3/2,2,5/2$ and $3$ to find the leading
corrections to the classical values.  By fitting the extrapolated GS
energy $e_0/s^2$ with a fitting function $f(s)=-1+a_1 s^{-a_2}$ we obtain a value for the exponent $a_2$ very close to one for
both reference states.
Hence, finally we have fitted the extrapolated CCM data for $e_0/s^2$
using  $f(s)=-1+x_1
s^{-1} + x_2s^{-2}$.  The fits yield the corresponding values $x_1=-0.414$ ($-0.410$) and $x_2=0.018$
($0.020$) for the $\sqrt{3}\times\sqrt{3}$ ($q=0$) reference states, as shown in Fig.~\ref{fig6}(a). 
Obviously, the $\frac{1}{s}$ prefactor is  close to that of the linear
spin-wave theory, and the contribution of the next order is small.
Note, however, that if the fitting function $f(s)$, with the above given values for $x_1$
and $x_2$, is applied to the $s=1/2$ case, it does not reproduce the GS selection of the $q=0$
state in this extreme quantum case [c.f., Fig.~\ref{fig6}(a) at 
the value $1/s = 2$].

Next we use $g(s)=1+b_1 s^{-b_2}$ as a fitting function to
fit the extrapolated CCM order parameters $M/s$, again using data 
for $s=3/2,2,5/2$ and $3$.
For the exponent $b_2$ we get the values 
$b_2=0.529$ for the $\sqrt{3}\times\sqrt{3}$ reference state 
and $b_2=0.666$ for the $q=0$ reference state. 
Clearly, the leading correction is not proportional to
$s^{-1}$; rather Chubukov's result\cite{chub92} of a 
leading correction proportional to $s^{-2/3}$
is confirmed for the $q=0$ state.
By contrast, for the
$\sqrt{3}\times\sqrt{3}$ reference state our results are in favor of
a leading correction for the order parameter 
proportional to $s^{-1/2}$.
Hence,  finally we have fitted the extrapolated CCM data for $M/s$  using
the fitting functions
$g(s)=1+y_1
s^{-1/2} + y_2s^{-1}$ [$g(s)=1+y_1
s^{-2/3} + y_2s^{-4/3}$] for the $\sqrt{3}\times\sqrt{3}$ [$q=0$] reference states.  The fits yield the corresponding values $y_1=-1.058$ [$-1.000$] and $y_2=-0.094$ [$0.006$], as shown in  
Fig.~\ref{fig6}(b).

\section{summary}
In the present investigation we present data for the GS energy per spin, $e_0$, and the order
parameter (sublattice magnetization), $M$,
of the kagome HAFM for spin quantum numbers $s=1/2,1,3/2,2,5/2,3$, 
using high-order CCM-SUB$n$-$n$ calculations based on the $\sqrt{3}\times\sqrt{3}$  and
the
$q=0$ reference states.
Our best estimate of the GS energy for the $s=1/2$ case, viz., $e_0=-0.4372$,  is clearly below rigorous upper bounds reported
recently,\cite{Evenbly2010,Yan2011} and it also agrees well  with recent accurate
DMRG\cite{Yan2011} and
ED\cite{lauchli2011} results.
We find that the GS selection by quantum fluctuations depends on the
spin quantum number $s$. While for $s=1/2$ the $q=0$ state is selected, for all values $s>1/2$
the  $\sqrt{3}\times\sqrt{3}$ state has lower energy.    
The order parameter  $M/s$ obtained by 
an appropriate extrapolation of the CCM-SUB$n$-$n$ data
vanishes for $s=1/2$ and $s=1$, but we get small (but nonzero) values $M/s>0$ for all values of the spin quantum number $s>1$.
Using CCM data for $s=3/2,2,5/2$ and $3$ we determine also the leading quantum corrections to the classical values of the GS energy and the order parameter. \\

\section*{Acknowledgments}
O.G. and J.R. thank R.~Zinke and J.~Schulenburg for fruitful discussions.

\end{document}